\documentclass[conference, letterpaper]{IEEEtran}
\IEEEoverridecommandlockouts
\usepackage{cite}
\usepackage[
    colorlinks=true,
    linkcolor=red,
    citecolor=orange,
    urlcolor=black,
    bookmarks=false
]{hyperref}
\usepackage{subcaption}
\usepackage{caption}
\usepackage{amsmath,amssymb,amsfonts,amsthm}
\usepackage{array}
\usepackage{algpseudocode} 
\usepackage{algorithm}
\usepackage{algorithmicx}
\usepackage{graphicx}
\usepackage{url}
\usepackage{bm}   
\usepackage{amsmath}
\usepackage{epstopdf}
\usepackage{verbatim}
\usepackage{xcolor}
\usepackage{enumitem}
\usepackage{wrapfig}
\usepackage{balance}
\usepackage{braket}
\usepackage{physics}
\usepackage{siunitx}
\usepackage{lipsum}  
\usepackage{tikz}
\usepackage{pgfplots}
\usepackage{mathtools}
\usepackage{siunitx}

\graphicspath{{figures/}}
\def\BibTeX{{\rm B\kern-.05em{\sc i\kern-.025em b}\kern-.08em
    T\kern-.1667em\lower.7ex\hbox{E}\kern-.125emX}}

\usepackage{listings}
\usepackage{color}

\definecolor{dkgreen}{rgb}{0,0.6,0}
\definecolor{gray}{rgb}{0.5,0.5,0.5}
\definecolor{mauve}{rgb}{0.58,0,0.82}
\begin{document}
\setlength{\columnsep}{0.22in}

\newtheorem{thm}{Theorem}
\newtheorem{lma}{Lemma}
\newtheorem{defi}{Definition}
\newtheorem{proper}{Property}

\title{Non-Abelian Mixer for QAOA on Hybrid Oscillator-Qubit Quantum Processors}

\author{
\IEEEauthorblockN {
   Thinh Le, 
   Hansika Weerasena, 
   Jianqing Liu
}

\IEEEauthorblockA {
  Department of Computer Science, North Carolina State University, Raleigh, USA 27606.\\
}
\IEEEauthorblockA {
    \{tvle2, hlokuka, jliu96\}@ncsu.edu
}

\thanks{This work is supported in part by the National Science Foundation under grants 2304118 and 2326746.}
}

\maketitle

\begin{abstract}
The realization of universal control in hybrid oscillator-qubit quantum processors enables the systematic design and implementation of quantum algorithms. However, the algorithmic development for such platforms remains at an early stage. While the Quantum Approximate Optimization Algorithm (QAOA) has been extensively studied in both continuous-variable (CV) and discrete-variable (DV) quantum systems, its development in the hybrid CV-DV setting remains limited. In this paper, we propose a hardware-native non-Abelian mixer for QAOA on hybrid CV-DV quantum processors and develop a corresponding hybrid ansatz for the Max-Cut problem. We evaluate the proposed ansatz on unweighted Erd\H{o}s--R\'enyi graphs and benchmark it against the standard transverse-field mixer using the approximation ratio and optimal-solution probability. Across all graph sizes and Fock cutoffs in our simulations, the proposed non-Abelian mixer consistently improves both expected solution quality and the probability of sampling an optimal solution relative to the transverse-field mixer. These results indicate that the proposed non-Abelian mixer is a promising building block for QAOA on hybrid oscillator–qubit platforms.

\end{abstract}
 
\thispagestyle{empty}
\IEEEdisplaynotcompsoctitleabstractindextext
\IEEEpeerreviewmaketitle

\section{Introduction}
\begin{figure}[t]
    \centering    
    \includegraphics[width=0.57\columnwidth]{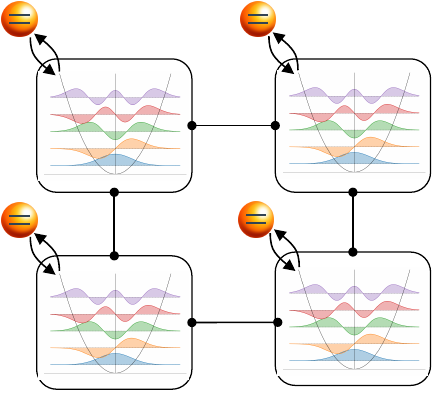}
    \caption{Schematic of a hybrid CV-DV quantum processor, illustrated with superconducting microwave resonators locally coupled to superconducting qubits and neighboring resonators coupled by beam splitters.}
    \label{fig:hybrid-QT}
\end{figure}

In the effort toward scalable quantum computing, many physical platforms, such as superconducting circuits~\cite{stancil2022principles}, trapped ions~\cite{bruzewicz2019trapped}, and neutral atoms~\cite{bohnmann2025bosonic}, have been explored as building blocks for quantum information processing. Across these platforms, quantum information can be encoded in either discrete-variable (DV) qubits or continuous-variable (CV) bosonic modes, each with its own advantages and limitations. There has been a growing attention to hybrid approaches that combine these two types of quantum degrees of freedom, where qubits and oscillators are used together as computational resources~\cite{liu2026hybrid}. Fig.~\ref{fig:hybrid-QT} illustrates one such hybrid CV-DV processor, consisting of superconducting microwave resonators working as CV bosonic oscillators, each coupled to a DV superconducting qubit. These hybrid CV-DV schemes are attractive because qubits offer effective control and readout, while oscillators provide an infinite-dimensional bosonic state space and natural support for bosonic operations. As a result, hybrid oscillator-qubit architectures have become an increasingly important setting for quantum control, quantum error correction, and algorithm design~\cite{ liu2025toward}.

Recent work has established systematic universal quantum control and measurement in such hybrid CV-DV systems ~\cite{liu2026hybrid, singh2025towards}. In particular, by utilizing the noncommutativity of the oscillator position $\hat{x}$ and momentum $\hat{p}$ quadratures, hybrid CV-DV systems inherently support non-Abelian control primitives, which provide a richer control algebra for oscillator–qubit operations. These primitives form the building blocks of non-Abelian quantum signal processing (NA-QSP)~\cite{liu2025toward},  with demonstrated applications including bosonic state preparation~\cite{hastrup2021measurement}, arbitrary qubit–oscillator state transfer, and the quantum Fourier transform~\cite{liu2025toward}. However, despite these initial demonstrations, the application of NA-QSP to broader classes of quantum algorithms remains open.


This motivates us to study the Quantum Approximate Optimization Algorithm (QAOA), which is a hybrid quantum-classical framework for finding near-optimal solutions to combinatorial optimization problems~\cite{farhi2014quantum}.  It operates iteratively by alternating between a cost Hamiltonian and a mixer Hamiltonian, with the goal of minimizing a given cost function. As a canonical NP-hard combinatorial optimization problem, the Max-Cut problem serves as a primary benchmark for QAOA. At the same time, Max-Cut is relevant to practical communication-network optimization, with formulations arising in interference-aware user grouping, binary link scheduling, and overlapping channel assignment in wireless networks~\cite{gu2024graph}. Efficiently solving Max-Cut is therefore important not only as a quantum-optimization benchmark, but also for communication applications. QAOA has been extensively studied in discrete-variable (DV) settings~\cite{wang2018quantum}, and continuous-variable (CV) formulations using bosonic modes have also been explored~\cite{enomoto2023continuous}. However, to the best of our knowledge, QAOA has not yet been studied on hybrid CV-DV processors. This also raises a key question: \textit{what is the right mixer for QAOA on a hybrid CV-DV processor?} To address this gap, we develop a hardware-native QAOA framework for hybrid CV-DV processors. Specifically, our paper makes the following contributions:
\begin{itemize}
    \item We propose a non-Abelian mixer for QAOA built from native instruction sets for hybrid CV-DV processors.
    \item We develop a hybrid CV-DV QAOA ansatz for the Max-Cut problem using the proposed non-Abelian mixer.
    \item We benchmark the proposed mixer against the standard transverse-field mixer and show consistent improvements across evaluation metrics.
\end{itemize}
In the rest of the paper, Section $\S$\ref{sec:preliminaries} introduces the hybrid CV-DV quantum computing, Max-Cut problem, and Gottesman-Kitaev-Preskill code. In Section $\S$\ref{sec:method} describes our proposed ansatz for Max-Cut QAOA using non-Abelian mixer. The simulation results and analysis are detailed in Section $\S$\ref{sec:sim}. Finally, Section $\S$\ref{sec:conclusion} summarizes our findings and outlines future research directions.

\section{Preliminaries}\label{sec:preliminaries}
In Section~$\S$\ref{subsec:hybrid-CV-DV}, we introduce universal control hybrid CV-DV systems. Then, Section~$\S$\ref{subsec:maxcut} formulates the Max-Cut problem and Section~$\S$\ref{subsec:GKP} presents the GKP code.
\subsection{Hybrid CV-DV Control Primitives}\label{subsec:hybrid-CV-DV}
The ability to perform universal control and measurement in hybrid CV-DV quantum systems is fundamental to the implementation of quantum algorithms and applications on such platforms. To make this control framework algorithmically useful, the underlying physical operations are exposed through an instruction set architecture (ISA), which provides the interface between hardware and algorithm design. 
Universal control of hybrid CV-DV quantum systems is commonly formulated through two primary ISAs: the phase-space ISA and the Fock-space ISA. The phase-space ISA relies on applying phase-space transformations that are dependent on the qubit state~\cite{eickbusch2022fast}, while the Fock-space ISA controls qubit evolution conditioned on discrete photon numbers~\cite{heeres2015cavity}. In this work, we focus on the phase-space ISA, in which each oscillator mode is described by the annihilation and creation operators $\hat{a}$ and $\hat{a}^\dagger$, respectively. Throughout, we represent the oscillator degree of freedom in terms of its two Wigner-unit quadrature operators, the position quadrature $\hat x$ and the momentum quadrature $\hat p$:

\begin{equation}
    \hat{x} = \frac{\hat a + \hat a^\dagger}{2}, \qquad
    \hat{p} = \frac{\hat a - \hat a^\dagger}{2i},
\end{equation}
which satisfy the canonical commutation relation $[\hat x,\hat p] = \frac{i}{2}$.

In the phase-space ISA, universal control is comprised of qubit rotations and qubit-controlled phase-space oscillator displacements, with beam-splitter gates included in the multimode setting. Local qubit control is described by the general single-qubit rotation:
\begin{equation}
    R_{\hat{n}}(\theta) = e^{-i\frac{\theta}{2} \hat{n} \cdot \vec{\sigma}},
\end{equation}
where $\theta$ denotes the rotation angle, $\hat{n} = \{n_x, n_y, n_z\}$ is a unit vector specifying the rotation axis, and $\vec{\sigma} = \{X,Y,Z\}$ is the vector of Pauli operators. The key primitive in the phase-space ISA is the conditional displacement (CD) gate:
\begin{equation}
    CD(\beta, \sigma_\phi) = e^{[(\beta\hat{a}^\dagger - \beta^*\hat{a}) \otimes \sigma_{\phi}]},
\end{equation}
where $\beta \in \mathbb{C}$ is the displacement amplitude, and $\sigma_{\phi} = \cos{\phi}~X + \sin{\phi}~Y$ denotes a Pauli operator in the $X$--$Y$ plane. This gate couples the CV and DV degrees of freedom by displacing the oscillator conditionally on the state of the ancilla qubit. Beyond the elementary phase-space ISA primitives, algorithmic constructions on the qubit register often require multi-qubit entangling gates. In this work, we focus on the logical $ZZ$ rotation gate:
\begin{equation}
    R_{ZZ}(\theta) = e^{-i \frac{\theta}{2}Z_1 \otimes Z_2},
\end{equation}
where $Z_1$ and $Z_2$ denote the Pauli-$Z$ operators acting on the first and second qubits, respectively. The $R_{ZZ}$ gate entangles two qubits by applying a parity-dependent phase shift in the computational basis. This gate can be synthesized using oscillator-mediated control by driving the shared oscillator along a closed phase-space trajectory, generating the desired parity-dependent entangling phase between the qubits~\cite{liu2026hybrid}.

\begin{figure*}[ht]
    \centering
    \includegraphics[width=0.60\linewidth]{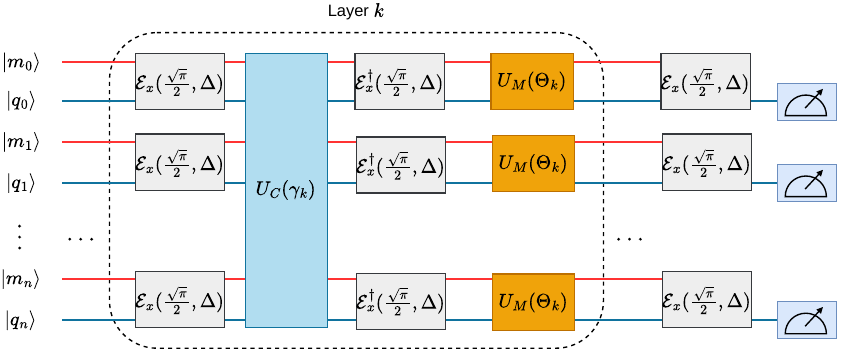}
    \caption{Hybrid CV-DV QAOA circuit for the proposed ansatz. In each layer $k$, $\mathcal{E}_x(\sqrt{\pi}/2,\Delta)$ couples the GKP-encoded oscillator modes to the ancilla register, the Max-Cut cost unitary $U_C(\gamma_k)$ applies graph-dependent phases, and \(\mathcal{E}_x^\dagger(\sqrt{\pi}/2,\Delta)\) transfers these phases back to the GKP encoding. The local mixers $U_M^{(i)}(\Theta_k)$ are then applied across oscillator--qubit pairs. A final $\mathcal{E}_x(\sqrt{\pi}/2,\Delta)$ is applied before qubit readout. Here, $m_i$ and $q_i$ denotes the $i$-th oscillator mode and ancilla qubit, respectively.}
    \label{fig:overview}
\end{figure*}

\subsection{The Max-Cut Problem}\label{subsec:maxcut}
We consider the standard QAOA benchmark problem of solving Max-Cut on unweighted Erd\H{o}s--R\'enyi random graphs \(G=(V,E)\) \cite{bollobas2011random}, where each edge is included independently with probability \(0.5\). Max-Cut is an NP-hard combinatorial optimization problem defined on graph $G$, where the goal is to partition the vertex set \(V\) into two subsets such that the number of edges between the two subsets is maximized. For an unweighted graph, this objective can be encoded by the cost Hamiltonian
\begin{equation}\label{eq:MaxCut-cost}
    H_C = \sum_{(i,j)\in E}\frac{1-Z_i Z_j}{2},
\end{equation}
where \(Z_i\) denotes the Pauli-\(Z\) operator acting on qubit \(i\). In this problem, the number of possible cuts scales exponentially with the number of nodes. As a baseline mixer, we consider the standard transverse-field mixer $H_M = \sum_{i=1}^n X_i$, where \(X_i\) denotes the Pauli-\(X\) operator on qubit \(i\)~\cite{farhi2014quantum}. The QAOA optimization objective is to maximize the expectation value $\langle H_C \rangle$ of the cost Hamiltonian.

\subsection{Gottesman-Kitaev-Preskill (GKP) Code}\label{subsec:GKP}
The GKP code \cite{liu2026hybrid} enables efficient quantum error correction by protecting logical information against small displacement errors in the phase space. In the ideal limit, the logical GKP codewords form infinitely sharp periodic combs in phase space, which are not normalizable. Physical realizations use finite-energy GKP states, where each ideal peak is broadened into a Gaussian and the overall comb is weighted by a Gaussian envelope. The finitely squeezed GKP logical states $\ket{0}_\mathrm{GKP}$ and $\ket{1}_\mathrm{GKP}$ are given by:
\begin{align}
    \ket{0}_\mathrm{GKP} = e^{-\Delta^2 \hat{n}}\sum_j \ket{\sqrt{\pi}(2j)}_x, \\
    \ket{1}_\mathrm{GKP} = e^{-\Delta^2 \hat{n}}\sum_j \ket{\sqrt{\pi}(2j+1)}_x.
\end{align}
where $\hat{n} = \hat{a}^\dagger\hat{a}$ represents the photon-number operator and the envelope parameter \(\Delta\) sets the width of the individual peaks. For finite-energy GKP codewords, the logical readout requires the ancilla qubit to rotate to orthogonal qubit states, depending on the encoded logical state. However, the position uncertainty of the Gaussian peaks induces qubit rotation errors, which can be corrected using NA-QSP. The non-Abelian QSP sequence for logical Z-basis readout of GKP codewords with envelope $\Delta$ is defined as~\cite{liu2026hybrid}: 
\begin{equation}\label{eq:logical_readout}
    \mathcal{E}_x\!\left(\frac{\sqrt{\pi}}{2},\Delta\right)
    =
    e^{\,i\sqrt{\pi}\hat{p}\Delta^2Y}\,
    e^{\,i\frac{\sqrt{\pi}}{2}\hat{x}X},
\end{equation}
where $X$ and $Y$ denote Pauli operators acting on ancilla qubit.

\section{Method}\label{sec:method}
At a high level, as illustrated in Fig.~\ref{fig:overview} for $n$ oscillator--qubit pairs, our hybrid QAOA ansatz represents the logical computational basis using finite-energy GKP codewords in the oscillator modes, while ancilla qubits mediate phase-space control and logical readout. In each QAOA layer, the operation $\mathcal{E}_x(\sqrt{\pi}/2, \Delta)$ coherently couples the GKP-encoded oscillators to the ancilla qubits, making the encoded logical state accessible to the qubit register for the Max-Cut cost unitary. The inverse operation $\mathcal{E}^\dagger_x(\sqrt{\pi}/2, \Delta)$ transfers the relative phases generated by $U_C(\gamma_k)$ onto the GKP-encoded logical state. The local non-Abelian mixer is then applied independently to each oscillator–qubit pair. The rest of this section first defines the proposed non-Abelian mixer and then describes the complete QAOA layer construction.
\subsection{Non-Abelian QSP Mixer}\label{subsec:na-Mixer}
Compared with Abelian QSP, which is restricted to commuting control variables, non-Abelian QSP exploits the canonical non-commutativity of quantum operators, e.g., the oscillator position $\hat{x}$ and momentum $\hat{p}$ quadratures to realize efficient universal control of hybrid CV-DV systems. The general non-Abelian QSP sequence is given as~\cite{singh2025towards}:
\begin{equation}
    U_{\vec\phi}(\hat{v}_1, \hat{v}_2, \ldots) = e^{i\phi_0\sigma_{\phi_0}}\prod_{j=1} \mathrm{CD}(\beta_j, \sigma_{\phi_j}),
    \label{eq:naqsp_cd}
\end{equation}
where $j$ indicates the conditional-displacement pulses in the NA-QSP sequence, $\hat{v}_j$'s are the phase-space quadratures, $\phi_0$ is the initial qubit phase, and $\beta_j$ is the j-th displacement amplitude.

In this work, we specialize Eq.~\eqref{eq:naqsp_cd} to a local two-generator ansatz acting on each oscillator-qubit pair. For QAOA layer \(k\), the global mixer is defined as
\begin{equation}
    U_M(\Theta_k)=\prod_{i=1}^{n} U_M^{(i)}(\Theta_k),
\end{equation}
where \(U_M^{(i)}(\Theta_k)\) is the local non-Abelian mixer that acts on the oscillator \(m_i\) and ancilla qubit \(q_i\). The parameter set \(\Theta_k\) is shared across all oscillator--qubit pairs within layer \(k\). The local non-Abelian mixer is defined as:
\begin{equation}
\label{eq:local_mixer}
\begin{aligned}
    U_M^{(i)}(\Theta_k)
    &=
    R_{X}(2\beta_0)\prod_{l=1}^{d}
    \Bigl[
        \mathrm{CD}(i\beta_x^{(l)},\sigma_{\phi_x^{(l)}})\,
        R_{X}(2\theta_x^{(l)}) \\
    &\qquad\qquad\quad
        \mathrm{CD}(\beta_p^{(l)},\sigma_{\phi_p^{(l)}})\,
        R_{X}(2\theta_p^{(l)})
    \Bigr],
\end{aligned}
\end{equation}
where $\Theta_k = \{\beta_0, \beta_x^{(l)}, \phi_x^{(l)}, \theta_x^{(l)}, \beta_p^{(l)}, \phi_p^{(l)}, \theta_p^{(l)}\}_{l=1}^d$ is the shared trainable parameter set, and $d$ is the mixer depth. We use phase-space primitives because GKP codewords are naturally defined by their periodic structure in oscillator phase space, and the phase-space ISA directly provides conditional displacements along the conjugate \(\hat{x}\) and \(\hat{p}\) quadratures. In Eq.~\eqref{eq:local_mixer}, the leading rotation \(R_{X}(2\beta_0)\) is an initial single-qubit rotation about the x-axis with trainable angle $\beta_0$. At depth $d=0$, the global mixer reduces to $U_M(\Theta_k) = \prod_{i=1}^n R_{X}(2\beta_0) = e^{-i\beta_0 \sum_i X_i}$, which is the standard transverse-field mixer. For $d \ge 1$, each block consists of two conditional displacements along the complementary $\hat{x}$ and $\hat{p}$ quadratures. The first gate, ~$CD(i\beta_x^{(l)},\sigma_{\phi_x^{(l)}})$, has a purely imaginary displacement amplitude, which couples the ancilla to the oscillator through the position quadrature $\hat{x}$. The second gate,~$CD(\beta_p^{(l)},\sigma_{\phi_p^{(l)}})$, has a purely real displacement amplitude and therefore couples the ancilla to the oscillator through the conjugate momentum quadrature $\hat{p}$. The interleaved single-qubit rotations $R_{X}(2\theta_x^{(l)})$ and $R_{X}(2\theta_p^{(l)})$ change the qubit basis between successive conditional displacements, introducing additional trainable parameters that increase the tunability of the local mixer.

\subsection{Hybrid CV-DV QAOA Ansatz for Max-Cut}\label{subsec:full-ansatz}
We consider a hybrid CV-DV system of $n$ oscillator-qubit pairs $\{(m_i,q_i)\}_{i=1}^n$, where each oscillator $m_i$ encodes the finite-energy GKP logical qubit, and each ancilla qubit $q_i$ serves as the ancillary subsystem for control and readout. The circuit is initialized in the product state:
\begin{equation}
    \ket{\psi_0} = \bigotimes_{i=1}^{n}\left(\ket{+}_{\mathrm{GKP}}^{(m_i)}\otimes\ket{0}^{(q_i)}\right),
\end{equation}
where $\ket{+}_{\mathrm{GKP}} = \frac{1}{\sqrt{2}}\left(\ket{0}_{\mathrm{GKP}}+\ket{1}_{\mathrm{GKP}}\right)$ denotes the finite-energy GKP logical plus state. As shown in Fig.~\ref{fig:overview}, each QAOA layer $k$ alternates between encoding and decoding the GKP logical information via $\mathcal{E}_x$ and $\mathcal{E}_x^{\dagger}$, applying the Max-Cut cost unitary $U_C(\gamma_k)$ on the qubit register, and mixing the hybrid state with the non-Abelian mixer $U_M(\Theta_k)$. First, $\mathcal{E}_x\!\left(\frac{\sqrt{\pi}}{2},\Delta\right)$ is applied independently to every oscillator-qubit pair. This operation rotates the ancilla qubit controlled by the oscillator position quadrature, mapping the logical information of the GKP codeword into the qubit register for the subsequent cost unitary $U_C(\gamma_k)$. In Eq.~\eqref{eq:logical_readout}, the factor $e^{\,i\frac{\sqrt{\pi}}{2}\hat{x}X}$ rotates the ancilla qubit according to the oscillator position eigenvalue. For finite-energy GKP states, the standard deviation $\Delta$ of each peak causes position uncertainty, which induces ancilla rotation errors. The prefactor $e^{i\sqrt{\pi}\hat{p}\Delta^2 Y}$ provides the corresponding non-Abelian precorrection. 

Subsequently, the Max-Cut cost unitary is applied to the ancilla register:
\begin{equation}
\begin{aligned}
     U_C(\gamma_k)=e^{-i\gamma_k H_C}
    =\prod_{(i,j)\in E} e^{-i\gamma_k(1-Z_iZ_j)/2} \\
    \propto \prod_{(i,j)\in E} R_{ZZ}^{(i,j)}(-\gamma_k),
\end{aligned}
\end{equation}
where \(H_C\) is defined in Eq.~\eqref{eq:MaxCut-cost}, \(\gamma_k\) is the variational parameter for the cost unitary in layer \(k\). The cost unitary $U_C(\gamma_k)$ can be implemented as a product of logical $ZZ$ rotations, up to a global phase. For each edge $(i,j)$, the gate $R_{ZZ}^{(i,j)}(-\gamma_k)$ applies a parity-dependent phase on the qubit register, with a cut edge ($Z_iZ_j=-1$) acquiring a phase $e^{-i\gamma_k/2}$ and an uncut edge ($Z_iZ_j=+1$) acquiring a phase $e^{i\gamma_k/2}$. Accumulated over all edges in $E$, these parity-dependent phases encode the Max-Cut objective directly into the relative phases of the qubit register amplitudes. Next, the adjoint unitary $\mathcal{E}_x^\dagger(\frac{\sqrt{\pi}}{2},\Delta)$ is applied to each oscillator-qubit pair to coherently invert the operation induced by $\mathcal{E}_x(\frac{\sqrt{\pi}}{2},\Delta)$. This operation transfers the relative phase accumulated under the cost unitary from the ancilla register back to the GKP logical state encoded in the oscillator, so that the non-Abelian mixer can operate directly on the oscillator phase-space degrees of freedom. In addition, applying $\mathcal{E}_x^\dagger$ ensures that the oscillator carries the logical information at the start of each subsequent layer, preserving the layered structure of the QAOA ansatz across all $P$ layers.

Finally, the local non-Abelian mixer $U_M^{(i)}(\Theta_k)$ is applied independently to the corresponding oscillator--qubit pair. The mixer drives transitions among the encoded logical states through the oscillator phase space, exploring the Max-Cut solution space beyond the dynamics generated by a qubit-only mixer, i.e, the transverse field mixer. Accordingly, one layer of the proposed hybrid QAOA circuit is described by
\begin{equation}
\begin{aligned}
    U_k
    &= U_M(\Theta_k)
    \left(\prod_{i=1}^{n}\mathcal{E}_x^{(i)\dagger}\!\left(\frac{\sqrt{\pi}}{2},\Delta\right)\right)
    U_C(\gamma_k) \\
    &\quad \times \left(\prod_{i=1}^{n}\mathcal{E}_x^{(i)}\!\left(\frac{\sqrt{\pi}}{2},\Delta\right)\right).
\end{aligned}
\end{equation}
After all $P$ layers, a final logical readout map $\mathcal{E}_x(\frac{\sqrt{\pi}}{2},\Delta)$ is applied to transfer GKP-encoded logical information from the oscillator back into the ancilla qubit register. The ancilla qubits are measured in the Z-basis, yielding a bitstring $b \in \{0,1\}^n$  that corresponds to a candidate bipartition of the graph $G$.

\section{Simulation results and discussion} \label{sec:sim}
In this section, we first describe the simulation setup in Section~$\S$~\ref{subsec:setup}, then compare the proposed mixer with the transverse-field mixer in Section.~$\S$\ref{subsec:compare}, and finally study the effects of mixer depth and the GKP envelope parameter $\Delta$ in Section~$\S$~\ref{subsec:effect}.
\begin{figure*}[ht]
    \centering
    \begin{subfigure}[t]{0.4\textwidth}
        \centering
        \includegraphics[width=\linewidth]{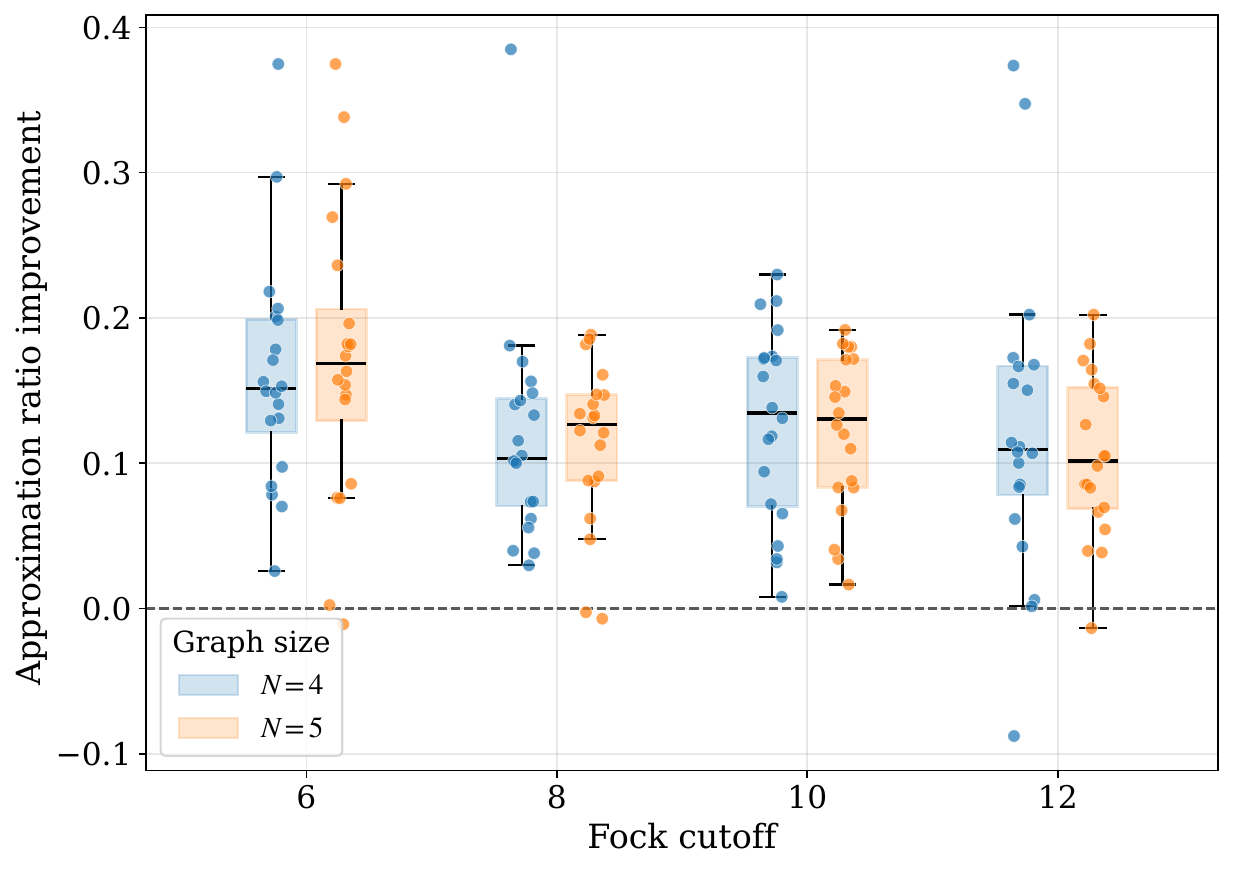}
        \caption{}
        \label{fig:exp1_ratio_delta}
    \end{subfigure}
    \hspace{2.5em}
    \begin{subfigure}[t]{0.4\textwidth}
        \centering
        \includegraphics[width=\linewidth]{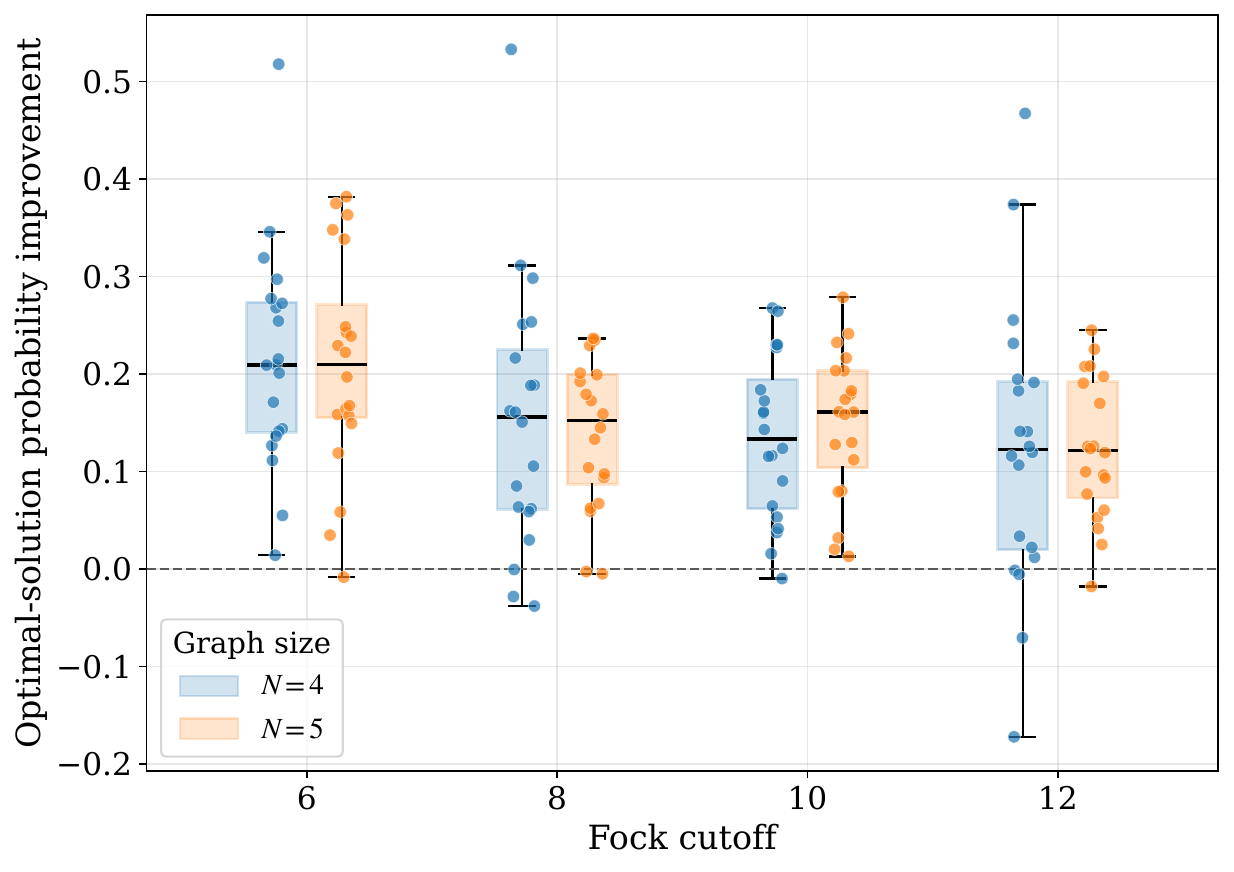}
        \caption{}
        \label{fig:exp1_popt_delta}
    \end{subfigure}
    
    \begin{subfigure}[t]{0.4\textwidth}
        \centering
        \includegraphics[width=\linewidth]{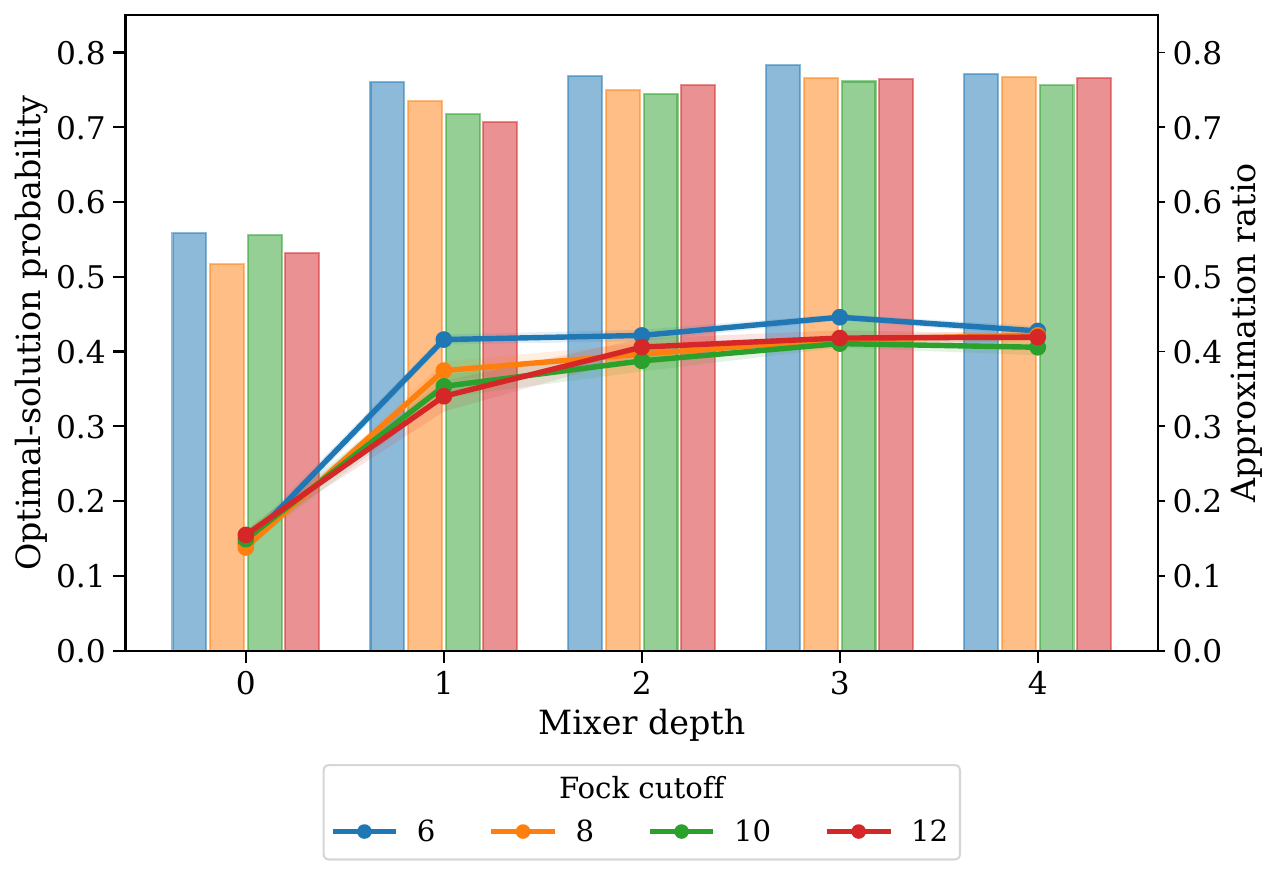}
        \caption{}
        \label{fig:exp2_popt_band}
    \end{subfigure}
    \hspace{2.5em}
    \begin{subfigure}[t]{0.4\textwidth}
        \centering
        \includegraphics[width=\linewidth]{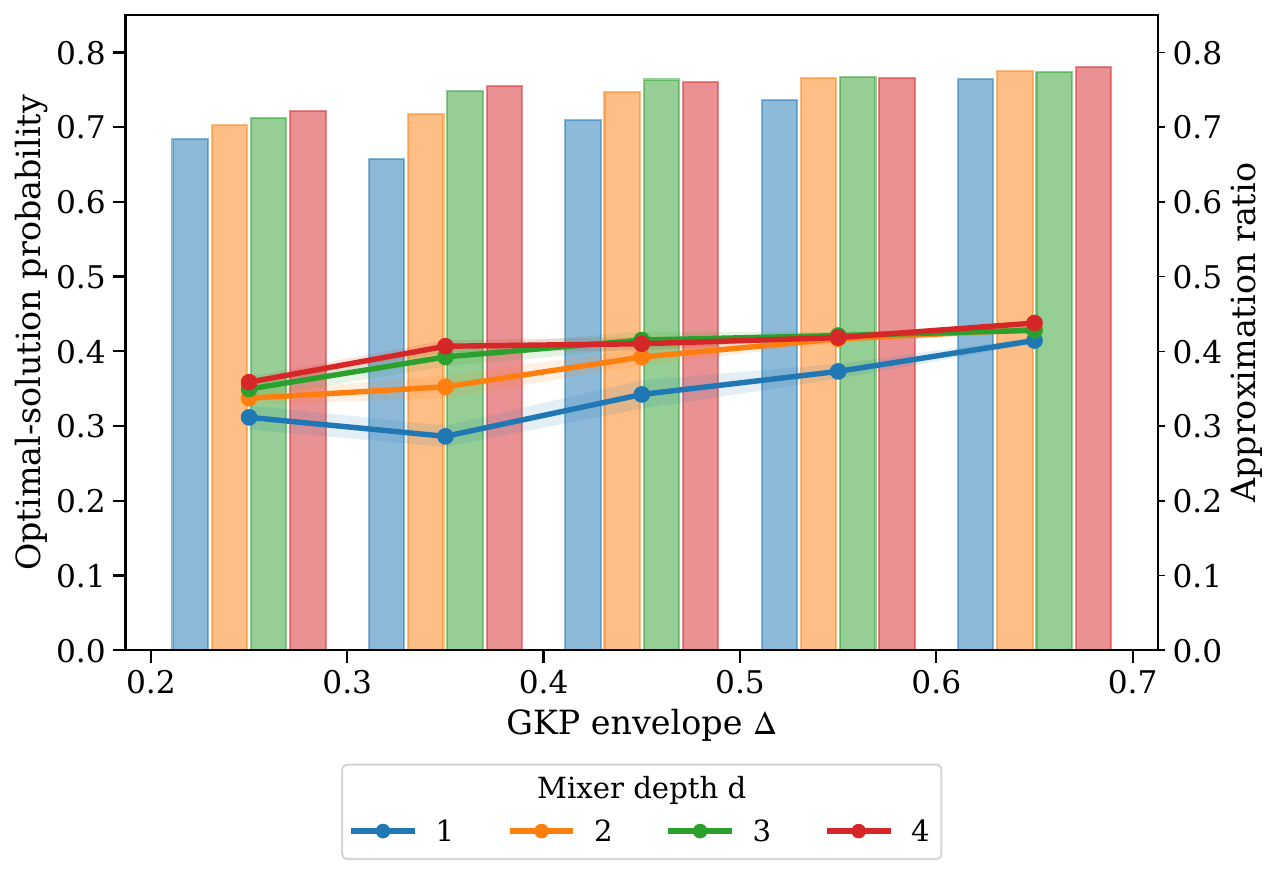}
        \caption{}
        \label{fig:exp2_ratio_band}
    \end{subfigure}
    \caption{
    Numerical results for the proposed non-Abelian mixer: (a) and (b) show the improvement of the proposed mixer over the standard transverse-field mixer across 20 random graph instances, measured in approximation ratio and optimal-solution probability, respectively. (c) shows the mean performance over repeated optimizations as a function of mixer depth \(d\) with \(N=4\) graph instance and \(\Delta=0.45\). (d) shows the corresponding mean performance as a function of \(\Delta\) at fixed \(N=4\) and \(N_{\max}=10\). In (c) and (d), bars denote optimal-solution probability and lines denote approximation ratio.
    }
    \label{fig:main_results}
\end{figure*}

\subsection{Simulation Setup}\label{subsec:setup}
The discussion presented in Section~$\S$\ref{sec:method} is formulated on infinite-dimensional oscillator modes, which are not physically realizable. Thus, in our simulations, each oscillator mode is truncated to the first $N_{\max}$ Fock states, giving an effective Hilbert-space dimension $N_\mathrm{max} + 1$. Here, Fock states are photon-number eigenstates of the oscillator. We evaluate the performance of the proposed hybrid CV-DV QAOA ansatz using metrics derived from the measured output bitstrings. 
Let \(z\in\{0,1\}^n\) denote an output bitstring with probability \(Pr(z)\), and let \(C(z)\) denote its cut value, i.e., the number of edges cut by the partition encoded by \(z\). The expected cut value can be written as:
\begin{equation}
    \langle H_C \rangle = \sum_{z\in \{0,1\}^n} Pr(z)\, C(z),
\end{equation}
Our primary performance metric is the approximation ratio, which is defined as the expected solution quality relative to the optimal solution:
\begin{equation}
    \mathrm{Approx.~Ratio} = \frac{\langle H_C \rangle}{\max_{z \in \{0,1\}^n} \langle z| H_C | z \rangle},
\end{equation}
In addition to approximation ratio, we also report the optimal-solution probability, which gives the probability of sampling an optimal bitstring. It is defined as:
\begin{equation}
    P_{\mathrm{opt}}=\sum_{z:\,C(z)=C_{\max}} Pr(z),
\end{equation}
where \(C_{\max}=\max_{z\in\{0,1\}^n} C(z)\). Thus, the approximation ratio measures expected solution quality relative to optimum, while \(P_{\mathrm{opt}}\) measures how often an optimal cut is sampled. 

Using these metrics, we first compare the proposed NA-QSP mixer against the standard transverse-field mixer on unweighted Erd\H{o}s--R\'enyi graphs with edge probability \(0.5\). We consider graph sizes $N\in\{4,5\}$, and Fock cutoffs $N_{\max}\in\{6,8,10,12\}$. In this comparison, the QAOA depth is fixed at $P=2$ and the NA-QSP mixer depth is fixed at $d=2$, with $20$ random graph instances generated for each $\{N, N_{\max}\}$ setting. Next, we fix a single $N=4$ graph instance to analyze the effect of mixer depth $d \in \{0,1,2,3,4\}$, and then fix the same graph together with $N_{\max} = 10$ to study the effect of the GKP envelope parameter $\Delta \in \{0.25, 0.35, 0.45,0.55,0.65\}$. In the latter two studies, each point is averaged over $10$ runs. In all simulations, conditional displacements are implemented with Z-basis qubit control, and variational parameters are optimized using a multistart COBYLA optimizer.

\subsection{Comparison with the transverse-field mixer}\label{subsec:compare}
Fig.~\ref{fig:exp1_ratio_delta} and~\ref{fig:exp1_popt_delta} compare the proposed NA-QSP mixer with the standard transverse-field mixer by reporting the improvement in each metric. Here, improvement refers to the difference between the corresponding metric obtained with the NA-QSP mixer and that obtained with the transverse-field mixer. The most important observation is that the NA-QSP mixer yields a consistently positive improvement in both approximation ratio and optimal-solution probability across all tested cutoffs and for both graph sizes $N=4$ and $N=5$. Averaged over all Fock cutoffs, the mean approximation-ratio improvement is approximately $0.132$ for $N=4$ and $0.128$ for $N=5$, while the corresponding mean improvements in optimal-solution probability are approximately $0.156$ and $0.155$, respectively. Moreover, the median improvement remains positive for every tested $(N,N_{\max})$ pair, indicating that the advantage persists across the sampled graph instances.

These results indicate that replacing the transverse-field mixer with the proposed NA-QSP mixer yields a more effective hybrid CV-DV QAOA ansatz for Max-Cut. In particular, the increase in optimal-solution probability shows that the advantage of the NA-QSP mixer is not only reflected in the expected value of the cost Hamiltonian, but also in the probability of sampling optimal bitstrings.

\subsection{Effect of mixer depth and GKP envelope parameter}\label{subsec:effect}
\textbf{Mixer depth dependence.} Fig.~\ref{fig:exp2_popt_band} shows the dependence of the optimal-solution probability and the approximation ratio on the NA-QSP mixer depth $d$ for a fixed $N=4$ graph instance. For optimal-solution probability, the most significant improvement is observed when the mixer depth increases from $d=0$ to $d=1$, where $d=0$ corresponds to the transverse-field baseline. This shows that the first NA-QSP mixer layer already yields an improvement in the optimal-solution probability. For example, at $N_{\max}=10$ the mean optimal-solution probability increases from approximately $0.149$ at $d=0$ to $0.353$ at $d=1$. As we add more layers to the mixer, performance improves marginally. At a fixed depth, the optimal-solution probability remains stable as $N_{\max}$ increases once the Fock cutoff is sufficiently large for the graph instance considered. The approximation ratio follows a similar trend. At $N_{\max}=10$, the mean approximation ratio increases from approximately $0.555$ at $d=0$ to $0.718$ at $d=1$. Taken together, for a graph size of $N=4$, these results suggest that the performance depends more on mixer depth than on the size of the truncated oscillator space.

Fig.~\ref{fig:exp2_ratio_band} demonstrates the dependence of the optimal-solution probability and the approximation ratio on the GKP envelope parameter \(\Delta\) at fixed cutoff $N_{\max}=10$. For mixer depths $d\geq1$, increasing $\Delta$ results in a slight improvement in the optimal-solution probability. At $\Delta=0.65$ and $d=4$, the mean optimal-solution probability reaches approximately $0.438$. The approximation ratio follows a similar trend, which also improves with increasing $\Delta$. For example, at mixer depth $d=1$, increasing $\Delta$ from $0.25$ to $0.65$ raises the mean approximation ratio from about $0.684$ to $0.765$. In finite-energy GKP constructions, $\Delta\to 0$ recovers the ideal codewords, corresponding to sharper peaks and higher-energy states. At fixed Fock cutoff, these sharper states are more difficult to represent in the truncated oscillator space. By contrast,  with $N_{\max}=10$, larger $\Delta$ produces approximate GKP states that are less susceptible to errors induced by Hilbert-space cutoff.

\section{Conclusion and Future Work}\label{sec:conclusion}
In this work, we take a step toward algorithm design for hybrid oscillator-qubit quantum processors. We show that the phase-space instruction set provides the native primitives for realizing QAOA in the hybrid CV-DV setting. Our proposed non-Abelian mixer, which is composed of these native phase-space primitives, consistently outperforms the standard transverse-field mixer in both optimal-solution probability and approximation ratio. At the algorithmic level, these results suggest that the proposed non-Abelian mixer enables more efficient exploration of the Max-Cut solution space. More generally, they highlight a broader design principle for hybrid CV-DV quantum platforms: algorithm design should follow the native hardware structure and control primitives, rather than rely on naive transfers of DV quantum algorithms.
For future work, an immediate next step is to study more expressive families of non-Abelian mixers, their trainability as circuit depth increases, and their behavior on larger problem instances and other combinatorial optimization tasks.

\bibliographystyle{IEEEtran}
\bibliography{refer}
\end{document}